# STUDY ON EMERGING APPLICATIONS ON DATA PLANE AND OPTIMIZATION POSSIBILITIES

Gereltsetseg Altangerel and Tejfel Máté

Department of Programming Languages and Compilers,
ELTE, Eötvös Loránd University, Budapest, Hungary

*ABSTRACT*

*By programming both the data plane and the control plane, network operators can adapt their networks to their needs. Thanks to research over the past decade, this concept has more formulized and more technologically feasible. However, since control plane programmability came first, it has already been successfully implemented in the real network and is beginning to pay off. Today, the data plane programmability is evolving very rapidly to reach this level, attracting the attention of researchers and developers: Designing data plane languages, application development on it, formulizing software switches and architecture that can run data plane codes and the applications, increasing performance of software switch, and so on. As the control plane and data plane become more open, many new innovations and technologies are emerging, but some experts warn that consumers may be confused as to which of the many technologies to choose. This is a testament to how much innovation is emerging in the network. This paper outlines some emerging applications on the data plane and offers opportunities for further improvement and optimization. Our observations show that most of the implementations are done in a test environment and have not been tested well enough in terms of performance, but there are many interesting works, for example, previous control plane solutions are being implemented in the data plane.*

*KEYWORDS*

*Data plane, load balancing, in-network caching, in-network computing, in-network data aggregation, INT*

## 1. INTRODUCTION

Software Defined Networking (SDN) is the beginning of deeply programmable network. Before SDN, network elements are only programmed by device vendors. SDN defines two important properties in network: first, the isolation of the data plane and the control plane functionalities, and second, the control plane on centralized controller(s) can control multiple network elements using a well-defined APIs (Application Programming Interface) such as OpenFlow. This not only simplifies network management, but also makes the network more open.

As a result of SDN's development over the last 10 years, we now have open platforms that can program every single element of the network. Moreover, network programming capabilities shifted from network device manufacturers to network operators, and new updates began to enter the network very quickly. In traditional network, it takes 4-7 years for a new technology to be approved by a standard organization and introduced into a network. Nowadays, such steps are not necessary, so many new technologies and innovations are developing rapidly in the network. However, some experts say the downside is that end users or companies are sometimes confused as to which of these many innovations to choose. This is a proof that the network is undergoing a lot of evolution and change.

Therefore, a fully programmable network has two pillars: a programmable data plane and a programmable control plane. The data plane programming began to be discussed in the 2000s with the advent of the merchant chip, but it became a reality in 2015, and research in this area is in great demand today [1]. The fundamental studies to make data plane programming more realistic are to develop a plane data programming language, create a programmable switching architecture, improve the performance of programmable switches and develop applications on data plane. This paper outlines applications on data plane and explores some improvement and optimizations on these applications.

The rest of this survey paper is organized as follows. Section 2 provides background information on data plane programming, Section 3 describes the data plane applications and future optimization ideas, and the final section presents conclusions.

## 2. BACKGROUND

### 2.1. Data plane programmability

Network devices use two kinds of algorithms to process packets: control plane and data plane. Data plane algorithms define the packet processing pipelines on device, while control plane algorithms define rules for manipulating a packet in the data plane, sense network, detect network failures, and update packet processing rules. In the SDN network, the control plane algorithms running on the controller platform (e.g., server) manage the data plane. For example, routing algorithms in the control plane define packet forwarding rules based on the destination IP address. These rules are installed in the routing table of the data plane via API. This means control plane and data plane are communicated using API. Programmable data planes are the latest concept in computer networking, and researchers are paying much attention to this area.

Thanks to the following concepts, technologies, and developments, a programmable data plane is becoming more practical.

1. **Data plane languages**: The key considerations to language design are to improve flexibility, support modular programming, and interface with other languages. Domain specific programming languages for defining data plane algorithms and functionalities (forwarding behaviour) are being developed. Examples include FAST[2], Domino[3], Protocol-Oblivious Forwarding[4], and NetKAT [5], P4 [6], with P4 being the most successful.
2. **Data plane architecture**: To map the data plane algorithm defined by domain-specific language to the switch ASIC hardware or software switch, the data plane architecture (programmable building blocks and data plane interface between them) must be provided by device vendors or language developers[7]. This architecture is also called a data plane model or hardware abstraction in some literature. For example, the P4 architecture team has recently developing a Portable Switch Architecture (PSA)[8] for the switches, and PNA Smart NICs [9].
3. **APIs**: Providing an interface for connecting the control plane and the programmable data plane. For example, the P4 compiler creates an API that connects the data plane to the control plane [7].
4. **Performance**: Thanks to numerous studies and technologies, the performance of a programmable switch has approached/same as that of a fixed-function switch [10].
5. **Applications:** Anyone, including universities, startups, and operators, can quickly develop more diverse application cases on a programmable data plane.

## 2.2. P4 Language

According to the annual P4 Workshop (2021), P4 has emerged as a leading domain-specific language for specifying programmable data planes, and thanks to strong communities and research, it has now a rich set of tools for P4 and a number of use cases. However, it still needs improvements: some vulnerabilities are identified by use cases and application development and are being fixed by the P4 communities. Also, additional features and capabilities are requested from industries. For example, Smart NICs have programmable packet processing pipelines and google is trying to implement congestion control mechanisms on Smart NIC and to implement it, they are requested some additional features from P4 community[11].

P4's advantages are target and protocol-independency. Target-independency means that P4 program can run on various targets such as hardware targets (switch and routers), and software targets such as BMv2 P4-OvS, and P4-DPDK[12], and number of supported targets are increasing. To ensure this feature, the hardware vendor or data plane language developers must define architecture and a compiler backend for a given target: provide them to the P4 developer [13] and so, the P4 program is easily mapped to the target with help of these. Protocol independency means that P4 developers can define their rich set of protocols and data plane behaviour/functionalities.

According to the architecture, P4 defines packet processing pipelines. The general architectures of the P4 data plane for research purposes are V1model and PSA [7]. Figure 1 describes basic pipelines in V1 model architecture including parser, match/action and deparper. Packet processing in this architecture is as follows: The parser receives incoming packets and extracts the header from them. Then, the match-action pipeline processes packet headers. A match-action block can contain one or more tables and actions.

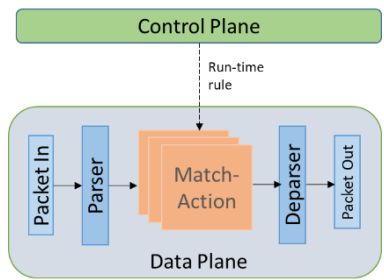
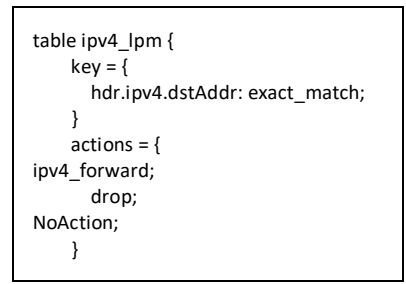

Figure 1. Abstract packet forwarding in P4        Figure 2. IPv4 table example

For example, the IPv4 routing table showed in Figure 2 can be created here, and the match key is the destination IP address and based on which, corresponding actions such as drop or forward are performed. In this block, the header can be added, subtracted, and modified. The deparser builds the outgoing packet by assembling the processed headers and the original payload [6]. In the case of other architectures, it is possible to define more detailed pipelines with more than one pair of parser and deparser for ingress and egress. Also, the match-action tables and external functions can be determined between parser and deparser.

## 3. APPLICATIONS ON THE DATA PLANE AND OPTIMIZATIONS

A programmable data plane allows anyone to quickly design, test, and deploy a variety of applications on a data plane. Currently, the focus is on the applications such as in-band telemetry, load-balancing, in-network computation, deployment of consensus protocols, congestion control, queue management and traffic management [14]. Since data centres are early adopters of the SDN network, most of these applications are currently designed for data centre networks. This section briefly describes some of these applications, their motivation, approach, challenges, and future improvement and optimization possibilities.

### 3.1. In-band Network Telemetry (INT)

The precise, fine-grained, real-time measurement and monitoring of the network is one of the foundations for organizing more optimal network. Traditional monitoring includes active methods (ping, traceroute), traffic mirror and SNMP-based inactive methods, and a combination of these [hybrid methods] [15]. They are simple to deploy, but the disadvantage is that, for example, passive methods have high processing costs, which can lead to performance limitations. In other words, the process of measuring the network itself can affect the state of the network.

In-Band Network Telemetry (INT) is a novel framework that collects network telemetry data such as the internal state of the switch from the data plane at line speeds. This kind of telemetry information is collected using a normal packet over a network or a probe packet (INT packet). Each node that receives INT packet embeds its metadata into this packet. Therefore, it does not put more strain on the network than traditional measurements. Also, it is more detailed, accurate, and close to real-time because it is implemented on data plane. One disadvantage is that metadata is limited by packet's maximum transmission unit (MTU). INT instructions (header) on what to collect from the devices are added to packet at the source INT node and then that packet is transmitted through network for collecting device's state. The metadata and INT header is removed from packet on the edge device (INT sink node). The sink node then performs the appropriate monitoring or actions, for example, it forwards the collected report to another external device or server for further monitoring [16]. Compared to other data plane applications, there are a large number of use cases, applications, and studies around INT. In the following part, we summarize some optimization works on those applications.

First of all, techniques for optimizing telemetry data include determining the proper size and structure of the INT packet within the MTU, minimizing the number of flows/packets for telemetry, and filtering unimportant telemetry information. For example, INT-label [17] is a lightweight network-wide telemetry architecture without explicitly using probe packets. This creates a sample packet by inserting a new field (instead of a protocol field) in the IP header. The sample packets are then sent periodically to collect device status. Therefore, it saves monitoring bandwidth and adapts seamlessly to topological changes. Preliminary evaluation on BMv2 software P4 switches shows that INT-label can achieve 98.54% network-wide visibility coverage under a label frequency of 100 times per second.

In addition, choosing optimal collection mechanisms is important, and event-based and policybased triggers are considered such mechanisms. For example, this solution [18] only triggers network monitoring if certain conditions are met. To achieve this, the Elastic Trie, a new data structure, has been introduced that allows the detection of (hierarchical) heavy hitters or super spreaders within the data plane. Implemented in P4, it is designed for two different FPGA target models and achieves high accuracy in detecting the targeted events with the memory constraints imposed by today's switches.

Moreover, optimizing the analysis process involves the use of advanced technology such as machine learning, simplifying the analysis process, reducing server load, monitoring server storage, and so on. A summary of these can be found in the survey paper [15].

## 3.2. In-Network Computing

Because traditional network devices are often designed to achieve high throughput, packet processing on data plane is limited. With programmable data plane, resource-intensive tasks can be run on network devices. This means that it is technically possible to offload a set of computing operations from an end-server and middlebox to a network device. Therefore, it might save energy for running servers, reduce traffic, thereby reducing network congestion, and process upper layer functionalities, such as transport and application layers at line speeds [19]. For example, Paxos is a consensus protocol for distributed networks in the application layer. It has been implemented in the data plane, and as a result, it improves the performance of the protocol itself and the performance of applications based on this protocol service [20].

Another segment that falls into this category is that it is now technically possible to offload most of the work on the control plane to the data plane: thus accelerating the control plane. For example, this work demonstrates the ability to offload basic tasks on the control plane, such as error detection, notification, connection recovery, and even a policy-based routing protocol into the data plane [21]. However, some tasks require a lot of resources and are not optimal for running on a data plane.

First of all, it is important to determine what type of computation operations or control plane tasks are most optimal to run on data plane. According to the application working group in the P4 community, the most feasible applications are in-network packet aggregation, in-network caching, and so on.

Second, thanks to the flexibility of programmability, network functionalities and applications can be chained in any order on data plane. But each one have its own requirement to work on the target: performance, computations ability, and so on. Therefore, it may not be possible to run them all on the same target due to different requirements. One interesting tool, Flightplan [22], is designed for disaggregating data plane programs and mapping them to suitable targets. This tool gives a possibility to run the programs in a distributed system that uses a variety of data plane targets that offer diverse computation and performance. The paper states that the idea has been implemented and achieved reasonable results.

### 3.2.1. In-Network Packet Aggregation

The search, query processing, dataflow computing, graph processing, and stream processing, and deep learning frameworks are the group of applications with partition/aggregation patterns in the data centre network. During the partitioning phase, job requests are subdivided into concurrent sub-tasks on different worker servers, and each worker server produces partial results. In order to obtain the final result, the partial results are collected and aggregated at the aggregation stage. In the aggregation phase, partial results must be passed between a large numbers of workers, which puts a significant load on the network. For example, Facebook's data trace shows that 46 percent of total traffic is generated during the aggregation phase. Furthermore, it leads to network congestion [23].

Therefore, traffic congestion can be reduced if the data aggregation function, which is usually performed on the application layer, is performed on a network path. Other reasons for in-network packet aggregation are that behind these functions are usually simple arithmetic logic operations, so placing them on a switch is simple, and there is no need to consider packet sequencing because these algorithms are associated associative functions.

SwitchML [24], using a programmable Tofino switch data plane, is designed to accelerate distributed parallel training in a machine learning model: reducing the volume of data exchanged from multiple workers on the network.

The next beneficial segment to use in-network aggregation is the integration of a small-sized and large number of packets. For example, merging IoT packets into a single transmission unit can reduce the additional computational costs associated with each transfer. Wang et al. [25] introduce proof-of-concept designs and implementations on IoT packet aggregation and disaggregation purely in P4 pipelines of the switching ASIC. The idea has been around for a long time, and now the opportunity to make it a reality may come with a programmable data plane.

### 3.2.2. In-Network Caching

You may need to access hundreds or thousands of storage servers in the background to view a single website. Understandably, there are a huge number of storage systems behind modern networking services, such as search engines, social networking and e-commerce, which are used by billions of users. An important way to improve storage system performance is to create a cache, where high-access items (hot items) are temporarily stored in the cache to allow users to access items on the storage system more quickly, and the cache must be constantly updated. One of the problems here is that hot items can change suddenly, which can lead to network flow imbalances when users start accessing those hot items in large quantities. For example, 60-70 percent of Facebook users access 10 percent of the total content [26]. Therefore, when building a caching system, these issues need to be considered.

Traditional networks use flash-based caches, disk-based caches, and server-based caches, and data plane programming provides new opportunity to create a cache on a programmable network device. Because network devices are naturally located on the path between the client and the server, creating a cache on it can further reduce latency.

The key-value storage data structure is often used to create a database in the cache. Netcache [27] is new key-value store architecture by leveraging flexibility, and the power of a modern programmable switch to handle queries on hot items of the storage server. It is built on top of rack (ToR) switch in the data centre network. Therefore, ToR switch plays important role and has 3 main modules: L2/L3 routing, on-path caching for key-value items, and query statistics. The Query statistic module identifies the hot items, and based on these statistics, the controller updates the cache. The core of Netcache is packet-processing pipeline which detect, index, store and serve key-value items. For example match-action table classify key on packet header and values are stored in register array, on-chip memory in programmable switch. One ToR switch can cache items on a storage server only connected to it, and cannot work with other ToR switches in a coherent way.

IncBricks [28] is another in-network, key-value store system built in a programmable data plane. What distinguishes it from Netcache is that it is implemented in the core, aggregation and ToR switch of the data centre network, as well as end-host server, and maintains the cache coherence using a directory-based cache coherence protocol.

These works are good start for in-network caching and both reduce latency by a certain percentage. The Netcache architecture was created on a Tofino and commodity server-based switch with a P4 pipeline, while IncCache was developed on Cavium XPliant switch and the forwarding plane was defined in a proprietary language.

According to the discussion on those works, the following questions can be open in the future: Mostly network requests (read) are processed from the cache. So, can write/delete requests be processed from cache? Do you need compression to reduce the cache size?, and so on.

### 3.3. Load Balancing Applications

The main purpose of a load balancer is to efficiently distribute the load to multiple parts of the network infrastructure to increase throughput, reduce response time, and prevent the overloading of a single resource. There are two main balancers: the layer 3 (L3) load balancer (s) selects one of the many routes that can direct the packet, while the layer 4 (L4) load balancer (s) chooses the one of serving instances (servers) for the incoming service request [29]. The data centre network has many redundant resources, so load balancers play an important role.

L3 load balancing mechanisms in the Data Centre network and Internet try to choose the congestion-free and optimal path from the multiple paths, so that bisection bandwidth can be used more efficiently. These mechanisms are usually implemented on the data plane. The most commonly used method is the Equal-Cost Multi-path Routing (ECMP), and because each flow is randomly allocated one of the same cost routes, performance may be reduced if elephant flows are allocated in the same path [30]. Another disadvantage is that ECMP does not track the overused path and ignores congestion (congestion-oblivious). Conga [31] has improved ECMP, a mechanism that detects congestion and maintains the congestion status of each path on the leaf/spine switch in the data centre network. However, this is not a scalable mechanism due to the limited memory of the leaf switch. Also, it is expensive to redesign because it uses custom hardware (chip architecture needs to be modified).

There are currently some ideas and implementations of the L3 load balancer on a programmable data plane: The improved ECMP [32] on P4 tries to make the ECMP path selection more flexible by slightly modifying path selecting procedure. It has been tested on NS2, which increases performance, but there are still design limitations. Another solution on P4 is W-ECMP [30] and the main feature is that it incorporates traffic congestion information into a normal flow, similar to P4's Inband Network Telemetry (INT) concept, which increases the update rate as the network load increases.

HULA [33], programmed in P4, is explicitly designed for the programmable switch architecture and it is scalable and congestion-aware. Conga centralizes the congestion track at one point (leaf switch), while HULA does it in a distributed manner. In addition, it can automatically detect network failures.

In this section, I will briefly describe some of the issues about L4 load balancer solutions on programmable data plane that are typically implemented on commodity servers. There are two main aspects to the design of an L4 load balancer. First, evenly distribute incoming connections through networks and servers. Second, providing per connection consistency (PCC): the ability to map packets belonging to the same connection to the same server, even if there are presence changes to the active servers and load balancers. But, meeting both these requirements at the same time has been an elusive goal [34].

It was not easy to ensure the PCC because the switch ASIC does not have enough memory to store a large number of connection states. However, the amount of memory is constantly increasing. SilkRoad [35] was proposed as a load balancer on a programmable switching ASIC and implemented using 400 lines of P4 code. The performance measurement on SilkRoad show that it can balance 10 million connections at line speed. SHELL [36] tried to implement a stateless load balancer on P4-NetFPGA programmable switch, and it is easier to deploy on a network device than a statefulsolution. Moreover, SHELL is application-agnostic and load-awareness.

The main advantage of implementing it on programmable switch is that there is no additional software load balancer in between application traffic and application server. This allows balancing load at line rate. Network traffic is constantly changing, so load balancing mechanisms need to be congestion-aware, dynamic, and with low latency. The results of empirical analysis of these implementations seem reasonable. In the future, the researchers can do an analytical study in terms of optimality and scalability on these in order to look for opportunities improving dynamic nature.

### 3.4. In-Network Congestion Control

Congestion control is a mechanism that controls the entry of data packets into the network, enabling better use of shared network infrastructure and avoiding congestive collapse. Different types of congestion control methods are used in the Internet and data center networks because these networks have different characteristics and requirements [37]. The Internet uses TCP-based end-to-end congestion control, while the data centre network initially used improved transport protocols such as DCTCP, but more recently RDMA over Converged Ethernet v2 [25] that use Ethernet flow control in switches. In addition, data centres have other architectures based on random path selection to reduce congestion. The problems with current solutions are that Ethernet-based versions do not provide low latency and high throughput at the same time, and the random selection of path-based solutions can cause serious congestion and transfer delays due to the elephant packets.

Data plane programming capabilities also allow the development of other advanced solutions. For example, NDP [38], the new congestion control architecture of the data centre network, claims to provide low latency and high throughput in all traffic conditions. To achieve this, they used some techniques including a modified switch queuing algorithm, together with per-packet multipath forwarding, and a novel transport protocol that takes advantage of these network mechanisms. But they reported some problems such as being moderately expensive in terms of CPU resources required from end systems. NDP was implemented in Linux hosts, in a software switch, in a hardware switch based on NetFPGA SUME, in P4, and in simulation.

QCN (Quantized Congestion Notification) is a type of Ethernet-based congestion control that uses a three-point algorithm architecture that sends congestion feedback from receiver to sender. P4QCN [39] is a flow-level, rate-based, network-assisted congestion control protocol, implemented in the P4. It extends the QCN protocol to IP-routed networks and uses a two-point (CP-RP) architecture, which reduces the end-to-end latency and the packet loss rate. Its performance is compared to other algorithms in the simulation network environment.

The main requirements for the congestion control mechanism are simplicity, dynamic nature, high performance, and fairness in data flows. A promising technology that can enable all of this could be a programmable data plane.

## 4. CONCLUSIONS

With the help of open data plane programmability, it is possible to redesign (some of) network applications on the data plane. The P4 application working group and other developers explored promising applications such as in-band network telemetry, traffic aggregation, caching, and load balancing, creating several applications cases in the experimental environment and few in the practical environment. A common problem with current solutions are that they have not been fully tested in the real world, have not been adequately analysed for performance, and only consider network layer optimization that is not harmonized with the upper layer.

Today, data centres are a major player in this area, and there are many implementations in the real data centre network. Therefore, there are many opportunities to develop new applications for other networks. For example, industrial networks and applications are unique and require strict requirements such as very low latency, almost zero losses, and high reliability. Here are some of the works around this network, and the authors say it has increased performance: moving time critical computations like event-detection to the field devices [40] and in-network traffic reduction method that filters out the unnecessary data traffic [41].

There are possibilities for chaining data plane functionalities (encryption/compression) or data plan applications in an optimal way: for example, advanced routing and congestion control based on INT monitoring.

Data plane programmability allows for tight integration between the application and the network but, the developers should always consider how network-level optimization affects the top level. When designing applications, the main considerations are network topology, device position, target capabilities, network policy, and application-specific requirements. Optimization in the application revolves around these topics. This study outlines emerging data plane applications and opportunities for further improvement and optimization.

## ACKNOWLEDGEMENTS

The research has been supported by the project "Application Domain Specific Highly Reliable IT Solutions" implemented with the support of the NRDI Fund of Hungary, financed under the Thematic Excellence Programme TKP2020-NKA-06 (National Challenges Sub programme) funding scheme.

## REFERENCES

[1] Nick McKeown, "Programmable Forwarding Planes Are Here To Stay," 2018.
[2] M. Moshref, A. Bhargava, A. Gupta, M. Yu, and R. Govindan, "Flow-level state transition as a new switch primitive for SDN," ACM SIGCOMM Comput. Commun. Rev., vol. 44, no. 4, pp. 377–378, 2015, doi: 10.1145/2740070.2631439.
[3] A. Sivaraman et al., "Packet transactions: High-level programming for line-rate switches," SIGCOMM 2016 - Proc. 2016 ACM Conf. Spec. Interes. Gr. Data Commun., pp. 15–28, 2016, doi: 10.1145/2934872.2934900.
[4] H. Song, "Protocol-oblivious forwarding: Unleash the power of SDN through a future-proof forwarding plane," HotSDN 2013 - Proc. 2013 ACM SIGCOMM Work. Hot Top. Softw. Defin. Netw., pp. 127–132, 2013, doi: 10.1145/2491185.2491190.
[5] C. J. Anderson et al., "NetkAT: Semantic foundations for networks," Conf. Rec. Annu. ACM Symp. Princ. Program. Lang., pp. 113–126, 2014, doi: 10.1145/2535838.2535862.
[6] P. Bosshart et al., "P4: Programming protocol-independent packet processors," Comput. Commun. Rev., vol. 44, no. 3, pp. 87–95, 2014, doi: 10.1145/2656877.2656890.

[7] The P4 Language Consortium, "P4 16 Language Specification version 1.2.2," pp. 1–163, 2020, [Online]. Available: http://p4.org.

[8] A. W. Group, "Naming conventions Packet paths PSA Data types," pp. 1–70, 2019.

[9] L. Consortium, "P4 Portable NIC Architecture ( PNA )," 2021.

[10] G. Antichi, T. Benson, N. Foster, F. M. V Ramos, and J. Sherry, "Programmable Network Data Planes (Dagstuhl Seminar 19141)," Dagstuhl Reports, vol. 9, no. 3, pp. 178–201, 2019, [Online]. Available: http://drops.dagstuhl.de/opus/volltexte/2019/11295.

[11] N. Dukkipati and K. Weitz, "Programmability in NICs for Congestion Control and Transport," 2021.

[12] The P4 Language Consortium, "P4 16 Language Specification version 1.2.1," pp. 1–163, 2020.

[13] P. Voros, D. Horpacsi, R. Kitlei, D. Lesko, M. Tejfel, and S. Laki, "T4P4S: A targetindependent compiler for protocol-independent packet processors," IEEE Int. Conf. High Perform. Switch. Routing, HPSR, vol. 2018–June, no. August, 2018, doi: 10.1109/HPSR.2018.8850752.

[14] R. Bifulco and G. Retvari, "A survey on the programmable data plane: Abstractions, architectures, and open problems," IEEE Int. Conf. High Perform. Switch. Routing, HPSR, vol. 2018–June, 2018, doi: 10.1109/HPSR.2018.8850761.

[15] L. Tan et al., "In-band Network Telemetry: A Survey," Comput. Networks, vol. 186, no. December 2020, 2021, doi: 10.1016/j.comnet.2020.107763.

[16] The P4.org Working Group, "In-band Network Telemetry ( INT ) Dataplane Specification," P4.org Appl. Work. Gr., pp. 1–42, 2020.

[17] T. Pan, E. Song, C. Jia, W. Cao, T. Huang, and B. Liu, "Lightweight network-wide telemetry without explicitly using probe packets," IEEE INFOCOM 2020 - IEEE Conf. Comput. Commun. Work. INFOCOM WKSHPS 2020, pp. 1354–1355, 2020, doi: 10.1109/INFOCOMWKSHPS50562.2020.9162684.

[18] J. Kučera, D. A. Popescu, H. Wang, A. Moore, J. Kořenek, and G. Antichi, "Enabling eventtriggered data plane monitoring," SOSR 2020 - Proc. 2020 Symp. SDN Res., pp. 14–26, 2020, doi: 10.1145/3373360.3380830.

[19] A. Sapio, I. Abdelaziz, A. Aldilaijan, M. Canini, and P. Kalnis, "In-network computation is a dumb idea whose time has come," HotNets 2017 - Proc. 16th ACM Work. Hot Top. Networks, pp. 150–156, 2017, doi: 10.1145/3152434.3152461.

[20] H. T. Dang and D. Sciascia, "NetPaxos : Consensus at Network Speed The Promise of Software Defined Networking."

[21] Edgar Costa Molero, "Hardware-Accelerated Network Control Planes Edgar," no. 04, p. 8– 11+16, 2015.

[22] N. Sultana et al., "Flightplan: Dataplane disaggregation and placement for P4 programs," Proc. 18th USENIX Symp. Networked Syst. Des. Implementation, NSDI 2021, pp. 571–584, 2021.

[23] P. Costa et al., "NetAgg : Using Middleboxes for Application-specific On-path Aggregation in Data Centres," pp. 249–261.

[24] A. Sapio et al., "Scaling distributed machine learning with in-network aggregation," Proc. 18th USENIX Symp. Networked Syst. Des. Implementation, NSDI 2021, pp. 785–802, 2021.

[25] S. Wang, C. Wu, Y. Lin, and C. Huang, "High-Speed Data-Plane Packet Aggregation and Disaggregation by P4 Switches," vol. 4, 2019.

[26] B. Atikoglu, Y. Xu, E. Frachtenberg, S. Jiang, and M. Paleczny, "Workload analysis of a largescale key-value store," Perform. Eval. Rev., vol. 40, no. 1 SPEC. ISS., pp. 53–64, 2012, doi: 10.1145/2254756.2254766.

[27] X. Jin et al., "NetCache : Balancing Key-Value Stores with Fast In-Network Caching," no. Figure 1, pp. 121–136.

[28] J. Nelson and L. Ceze, "IncBricks : Toward In-Network Computation with an In-Network Cache."

[29] A. Aghdai, C. Y. Chu, Y. Xu, D. Dai, J. Xu, and J. Chao, "Spotlight: Scalable Transport Layer Load Balancing for Data Center Networks," IEEE Trans. Cloud Comput., 2020, doi: 10.1109/TCC.2020.3024834.

[30] J. L. Ye, C. Chen, and Y. Huang Chu, "A Weighted ECMP Load Balancing Scheme for Data Centers Using P4 Switches," Proc. 2018 IEEE 7th Int. Conf. Cloud Networking, CloudNet 2018, pp. 1–4, 2018, doi: 10.1109/CloudNet.2018.8549549.

[31] M. Alizadeh et al., "CONGA: Distributed congestion-aware load balancing for datacenters," Comput. Commun. Rev., vol. 44, no. 4, pp. 503–514, 2015, doi: 10.1145/2619239.2626316.

[32] L.-Q. HE and L. OU, "Research on Programmable Data Plane Load Balancing based on Multipath Selection," vol. 131, no. Eeeis, pp. 260–268, 2017, doi: 10.2991/eeeis-17.2017.36.
[33] N. Katta, M. Hira, C. Kim, A. Sivaraman, and J. Rexford, "HULA : Scalable Load Balancing Using Programmable Data Planes," 2016.
[34] T. Barbette et al., "A high-speed load-balancer design with guaranteed per-connectionconsistency," Proc. 17th USENIX Symp. Networked Syst. Des. Implementation, NSDI 2020, pp. 667–683, 2020.
[35] R. Miao, H. Zeng, C. Kim, J. Lee, and M. Yu, "Silkroad: Making stateful layer-4 load balancing fast and cheap using switching asics," SIGCOMM 2017 - Proc. 2017 Conf. ACM Spec. Interes. Gr. Data Commun., pp. 15–28, 2017, doi: 10.1145/3098822.3098824.
[36] B. Pit-Claudel, Y. Desmouceaux, P. Pfister, M. Townsley, and T. Clausen, "Stateless LoadAware Load Balancing in P4," Proc. - Int. Conf. Netw. Protoc. ICNP, vol. 2018–September, pp. 418–423, 2018, doi: 10.1109/ICNP.2018.00058.
[37] D. A. Alwahab and S. Laki, "A simulation-based survey of active queue management algorithms," ACM Int. Conf. Proceeding Ser., no. October, pp. 71–77, 2018, doi: 10.1145/3193092.3193106.
[38] M. Handley et al., "Re-architecting datacenter networks and stacks for low latency and high performance," SIGCOMM 2017 - Proc. 2017 Conf. ACM Spec. Interes. Gr. Data Commun., pp. 29–42, 2017, doi: 10.1145/3098822.3098825.
[39] J. Geng, J. Yan, and Y. Zhang, "P4QCN: Congestion control using P4-capable device in data center networks," Electron., vol. 8, no. 3, 2019, doi: 10.3390/electronics8030280.
[40] J. Vestin, A. Kassler, S. Laki, and G. Pongracz, "Toward In-Network Event Detection and Filtering for Publish/Subscribe Communication Using Programmable Data Planes," IEEE Trans. Netw. Serv. Manag., vol. 18, no. 1, pp. 415–428, 2021, doi: 10.1109/TNSM.2020.3040011.
[41] Györgyi, Kecskeméti, Vörös, Szabó, and Laki, "In-network Solution for Network Traffic Reduction in Industrial Data Communication," 2021, doi: 10.1109/NetSoft51509.2021.9492564.

## AUTHORS

**Tejfel Máté** received his B.Sc., M.Sc. and Ph.D. Degree in Computer Science, from ELTE, Budapest Hungary. He is currently working as an Associate Professor in the Department of Programming Languages and Compilers, ELTE. His research interest includes programming languages, correctness check, Software Defined Networks, and network optimization. For more information, visit his database at 0000-0001-8982-1398 orchid-id.

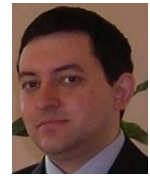

**Altangerel Gereltsetseg** received her B.Sc. & M.Sc. Degree in Information technology, from Mongolian University of Science and Technology (MUST). She is currently a Ph.D. student in the Department of Programming languages and compilers, ELTE under the supervision of Professor Tejfel Máté. Her research interest includes Software-defined Networks, deeply programmable network, and network optimization. For more information, visit her database at 0000-0002-1594-8158 orchid-id.

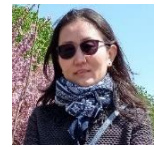